\documentclass[aps,prd,a4paper,dlspace,twocolumn]{revtex4-1}
\usepackage{color}
\usepackage{epsfig}
\newcommand{\be}{\begin{equation}}
\newcommand{\ee}{\end{equation}}
\newcommand{\ba}{\begin{eqnarray}}
\newcommand{\ea}{\end{eqnarray}}
\newcommand{\nn}{\nonumber\\}
\begin{document}
\title{On the Thermodynamic Geometry of Hot QCD}
\author{Stefano Bellucci$^{a}$}
\email{bellucci@lnf.infn.it}
\author{Vinod Chandra$^{b,c}$}
\email{joshi.vinod@gmail.com, Current affiliation of the author is  $c$}
\author{Bhupendra Nath Tiwari$^{a}$}
\email{tiwari@lnf.infn.it} \affiliation{$^{a}$ INFN-Laboratori
Nazionali di Frascati, Via E. Fermi 40, 00044 Frascati, Italy.}
\affiliation{$^{b}$ Department of Physics, Indian Institute of
Technology Kanpur, Kanpur- 208 016, India.}
\affiliation{$^{c}$ Department of Theoretical Physics, Tata Institute of Fundamental Research, Homi Bhabha Road
Mumbai-400005, India.}
\begin{abstract}
We study the nature of the covariant thermodynamic geometry
arising from the free energy of hot QCD. We systematically analyze
the underlying equilibrium thermodynamic configurations of the
free energy of 2- and 3-flavor hot QCD with or without including
thermal fluctuations in the neighborhood of the QCD transition
temperature. We show that there exists a well-defined thermodynamic
geometric notion for the QCD thermodynamics. The geometry thus
obtained has no singularity as an intrinsic Riemannian manifold.
We further show that there is a close connection of this geometric
approach with the existing studies of correlations and quark number
susceptibilities.

\vspace{2mm}
\end{abstract}
\maketitle

\section{Introduction}

Motivated from string theory, D-brane physics and black hole
thermodynamics, we focus on investigating the possibilities of
covariant thermodynamic geometric study of hot QCD near the QCD
transition temperature. In particular, we intend to study the
thermodynamic geometry arising from the free energy for the 2- and
3-flavor finite temperature QCD. The main idea of the present paper
is first to develop a geometric notion for QCD thermodynamics and then
to relate it with the existing microscopic quantities already known
from the physics of hot QCD \cite{hep-ph/0110369v3, hep-lat/0110032}.
We further incorporate thermal fluctuations in the free energy \cite{SM} and
thus study their contributions to the thermodynamic geometry.

Let us recall that the existence of the covariant geometric structure
in equilibrium thermodynamics was introduced by Weinhold \cite {wein}
through an inner product in the space of equilibrium thermodynamic
macrostates defined by the minima of the internal energy function
$U=U(\nu_i, v, s)$ as the Hessian function $ h_{ij}= \partial_i
\partial_j U $. Here, the reduced quantities $ \lbrace \nu_i, v, s
\rbrace $ may respectively be defined as the underlying chemical
potentials $\mu_i$, volume $V$ and that of the entropy $S$ scaled
with temperature of the considered equilibrium configuration. In
order to provide a physical scale and to restrict negative eigenvalues
of the metric tensor, we require that the volume and other parameters
be held fixed for the subsequent analysis of the QCD configurations.

Interestingly, the Riemannian geometric structure thus involved
indicates certain physical significance ascribed to hot QCD. The
associated inner product structure on the intrinsic space may
either be formulated in the entropy representation, as a negative
Hessian matrix of the entropy with respect to the extensive
variables, or that of the energy representation, as the Hessian
matrix of the associated energy with respect to the intensive
variables. Such geometric considerations have been studied earlier
in the literature as the covariant thermodynamics whose
application in the energy representation has first been considered
by Weinhold \cite {wein}. On the other hand,  the entropy
representation has been considered in Ruppeiner's fluctuation
theory \cite {rup1, rup2}.

In this paper, we shall take the former picture and study the
covariant thermodynamic geometric properties ascribed to hot QCD.
The thermodynamic macrostates of an underlying equilibrium
chemical potential space may in turn be described by the minima of
the energy function $F =F(U, V, N)$. Explicitly, the thermodynamic
metric tensor of an intrinsic Riemannian space spanned by the
chemical potentials may be given as $g_{ij}= \partial_i
\partial_j F( \mu_i, T, V, S)$ and turn out to be conformal to the
Ruppenier metric with the temperature as the conformal factor.
The intrinsic metric as an inner product on the (hyper-)surface
of chemical potentials is thus necessarily ensured to be
symmetric, positive definite and satisfies the triangle inequality
since the energy function has a minimum configuration in the
equilibrium.

On other hand, the thermodynamic geometry of the equilibrium
configurations thus described has extensively been applied to
study the thermodynamics of a class of black hole configurations
\cite{bnt}. Recent studies of the thermodynamics of diverse black
holes in the state-space geometric framework have elucidated
interesting aspects of phase transitions and their relations
associated with the extremal black hole solutions in the context
of the moduli spaces of $ N \geq 2 $ supergravity
compactifications \cite{Bull}. It may be argued however that the
connection of such geometric formulations to the fluctuation
theory of black holes thermodynamics requires several
modifications \cite {rup3}. The geometric formulation involved has
first been applied to $ N\geq 2$ supergravity extremal black holes
in $D=4$ which arise as low energy effective field theories from
compactification of Type II string theories on Calabi-Yau
manifolds \cite {fgk}. Since then, several authors have attempted
to understand this connection \cite{cai1, aman} both for
supersymmetric as well as non-supersymmetric black holes and five
dimensional rotating black rings. The geometric notions remain
well-defined for the diverse cases of extremal and non-extremal
black branes \cite{bnt}.

Here, we shall study the standard two dimensional thermodynamic
geometry associated with the space of chemical potentials $
\lbrace \mu_1, \mu_2 \rbrace $ in the framework of quasi-particle
theories of QCD. The application of the above geometric notions, as in
any conventional thermodynamical systems, suggests that the space
defined by the chemical potentials indicates the nature of the
underlying statistical system. In particular, such geometric
notions, like the covariant nature of the metric tensor and
non-zero scalar curvature, shed light on the interactions present
in the system. It is shown further that the associated conformally
related thermodynamic scalar curvature defines the chemical
correlation length of the system.

As a matter of fact, a few simple manipulations illustrate that the conformal thermodynamic curvature
is inversely proportional to the singular part of the free energy associated with long range correlation(s)
whose divergences consequently signify certain phase transition(s) at the critical point(s) of the QCD.
It is worth to mention that the Ruppeiner formalism may further be applied to diverse condensed matter systems
with two dimensional intrinsic Riemannian spaces, which has been found to be completely consistent with the scaling
and hyper scaling relations involving certain critical phenomena and in turn reproduces the corresponding
critical indices \cite{rup1}. 

The purpose of the present paper is to investigate the relation between
the covariant thermodynamic geometry and the quark-number
susceptibility tensor, both of which have recently received
considerable attention from the perspective of black hole physics
and QCD thermodynamics. Thus, the goal here is to explore
underlying microscopic arguments behind the notion of the
covariant thermodynamic geometry arising from the free energy of
certain QCD thermodynamic configurations. In general, it is however
known in the context of quasi-particle models \cite{hep-ph/0110369v3}
that the quark number susceptibilities are defined as $ \chi_{ij} \equiv
\frac{\partial \mathcal N_i}{\partial \mu_j}= \frac{\partial^2
P}{\partial \mu_i \partial \mu_j}= \chi_{ji} $. Here $i,\,j$ are
flavor indices, ${\mathcal N}_i $ is the quark number density, and
$P$ is the associated pressure of the QCD configurations.

In this framework, it has further been shown that lattice
measurements of the off-diagonal susceptibility $ \chi_{ud} $
determine the associated physical properties of the quark-gluon
plasma. As reported in Ref. \cite{hep-lat/0110032}, these are
shown to be consistent with Hard Thermal Loop (HTL) studies down
to the temperature of order as low as that of $ 2.5 T_{C} $, see
also \cite{hep-ph/0101103}. Blaizot, Iancu and Rebhan have in
particular analytically computed the quark number susceptibilities
for the quark gluon plasma at finite temperature within the
self-consistent resummation of Hard Thermal/Dense Loop (HTL/HDL)
\cite{hep-ph/0105183}. Interestingly, these predictions at zero
chemical potential were found to be consistent with the lattice
results.

Note that all the diagonal and off-diagonal elements of $\chi_{ij}$
become equal at $\mu_i= 0$, and thus the quark number susceptibility
tensor reduces to two different sectors, and in particular, we have
$ \chi_{ij}\Big|_{\mu=0} \equiv \chi $ for $i=j$, and that of
$ \chi_{ij}\Big|_{\mu=0} \equiv \tilde{\chi} $ for $i\not=j$.
In this paper, we have shown that the same remains true for the
components of the covariant thermodynamic metric tensor which may
further be regarded as the above quark number susceptibility tensor.
In turn, this equivalence explains that the involved macroscopic/
microscopic duality relations, in our investigations of the 2 and 3-flavor
QCD near transition temperature, turn out to be in a close connection
with the AdS/CFT correspondence \cite{AdS/CFT}. The present analysis
indicates a sort of duality relationship between the macroscopic
thermodynamic geometry and microscopic quark number susceptibility.
The above formalism remains valid away from the limit of $\mu_i \neq 0$.
We note further that the microscopic formalism is based on the
evaluation of the diagonal susceptibility $\chi $ in a chosen
resummation scheme, which directly follows from the expression
of the fermion number density $ \mathcal N $. From the viewpoints
of microscopic theory, it is worth mentioning that the susceptibility
tensor $\chi $ may be expressed in terms of the associated dressed
fermion propagators as a function of the self-energies and fermions
helicities \cite{ 065003}.

On the other hand, the off-diagonal susceptibility tensors may be
classified by the symmetry of the charge conjugation, or that of
the C-parity. In particular, this in terms of chemical potentials
means that a quark loop with two gluon external lines is
necessarily even in the chemical potentials, while a quark loop
with three gluon legs may also generate a linear term in the
chemical potentials, which turns out to be symmetric in the color
indices. We thus find an interesting match that the even derivative
geometric quantities like the components of the metric tensor or
the associated determinants and the thermodynamic scalar curvature
as the fourth derivative invariant quantity turn out to be even in
the chemical potentials. We envisage further that the components of
thermodynamic metric tensor microscopically correspond to those of
the susceptibility tensor and thus require two fermion loops
connected by a set of gluon lines.

As the final exercise, we discuss the thermodynamic geometry
of hot QCD including the thermal fluctuations over an equilibrium
configuration of the chemical potentials. As it is well known, any
thermodynamical system which may be considered as an ensemble has
logarithmic and polynomial contributions to the free energy
\cite{SM}. Moreover, such a connection have earlier been explored
from the perspective of black hole physics, showing that the
thermodynamic Ruppeiner geometry of the non-extremal rotating BTZ
black holes, as well as that of the BTZ-Chern Simons black holes,
get modified under the thermal fluctuations \cite{sgt}.  Subsequently,
the thermodynamic geometry remains flat with and without the higher
derivative contributions to the entropy.

In the context of QCD, what follows next that such geometric notions
implicitly assume an equilibrium statistical basis defined in terms
of an ensemble, and thus one may explore the role of thermal
fluctuations in the free energy and consequently on an associated
intrinsic Riemannian geometry corresponding to the quasi-particle
theories. For example, the canonical free energy corrected by the
contributions coming from thermal fluctuations may be regarded
as a closer approximation to that of the free energy in the
corresponding ensemble. We here show that such considerations
apply as well to the hot QCD ensembles, and the specific forms
of the logarithmic and polynomial contributions can easily be
calculated for a wide class of quasi-particles theories.
It is to be noted that the applicability of this analysis
presupposes that the underlying ensemble is thermodynamically
stable, which requires a positive specific heat, or
correspondingly that the Hessian matrix of the free energy
function must be positive definite. The free energy function for
any thermodynamical system incorporating such contributions
\cite{SM} can easily be written as $ F = F_0+ \ln (CT^2) $. In
this article, we have shown that such logarithmic contributions
to the free energy induces a set of bumps in the scalar curvature
of the intrinsic thermodynamic space, spanned by the quark chemical
potentials.

The paper is organized as follows. In the first section, we have
presented the motivations to study the intrinsic Riemannian space
obtained from the Gaussian fluctuations of the QCD free energy. In
particular, we have outlined microscopic implications of the
geometric QCD from the viewpoint of AdS/CFT correspondence. The
remaining significance of the present analysis are summarized in
the final section. In section 2, we have introduced very briefly
the notion of thermodynamic geometry for the quasi-particles.
In section 3, we obtain the 2- and 3-flavor QCD free energy and
temperature near the $T_C^0$, and express them as a function of
the quark chemical potentials and thereby incorporate the thermal
contributions into the free energy. In section 4, we analyze the
thermodynamic geometry of two-flavor QCD in the framework of the
quasi-particle theory with and without the logarithmic contributions.

In section 5, we focus our attention on the thermodynamic geometry
of the 3-flavor QCD configurations. Explicitly, we investigate the
nature of the thermodynamic geometry thus defined as an intrinsic
Riemannian manifold obtained from the free energies with or
without the contributions being considered in the section 4. We
have explained that such a geometry obtained from the free energy
of the quasi-particle theories results to be well-defined and
pertains to an interacting statistical system. Finally, section 6
contains concluding issues and discussion of the thermodynamic geometry
for both the 2-flavor and the 3-flavor hot QCD and the concerned
microscopic implications arising from the underlying QCD, and thus
it offers a physical explanation of finite chemical potential
macroscopic/microscopic duality relations in a close connection
with the AdS/CFT correspondence.

\section{Thermodynamic Geometry}

In this section, we present a brief review of the essential
features of thermodynamic geometries from the perspective of the
application to the hot QCD. In particular, we shall focus our
attention on the geometric nature of the QCD thermodynamics in the
neighborhood of QCD transition temperature, $T_C$ with a finite
number of quark chemical potentials introduced within the
framework of quasi-particles theories.

Let us consider an intrinsic Riemannian geometric model whose
covariant metric tensor may be defined as the Hessian matrix of
the free energy with respect to an arbitrary finite number of
chemical potentials carried by the quasi-particles in a given
thermodynamic configuration at a fixed volume or any other
parameters of the equilibrium QCD. In particular, let us define a
representation $ F(\mu_i, T) $ for given any free energy, chemical
potentials, and temperature $ \lbrace F, \mu_i, T \rbrace $ and
then choose a particular temperature slice as an intersection of
the line $ T:= T_C^{(0)} $ with the space spanned by the
underlying chemical potentials $ \lbrace \mu_i, T \rbrace $, that
is to define the temperature as in Eq. \ref{eq11}. In this
consideration, the free energy thus defined may further be shown
to be only a function of the quark chemical potentials, which in
fact we have indicated in general by Eqs. \ref{eq12}, \ref{eq14}.
This method in general turns out to be an intrinsic Riemannian
geometry, which is solely based on nothing other than the quark
chemical potentials. This thus yields that the space spanned by
the $n$ chemical potentials of the theory under consideration
exhibits a n-dimensional intrinsic Riemannian manifold $ M_n $.
The components of the covariant metric tensor, i.e. the so called
thermodynamic geometry \cite{rup1, rup2, sgt, aman} may be defined
as \be \label{eq1}
 g_{ij}:=\frac{\partial^2 F(\vec{x})}{\partial x^j \partial x^i},
\ee
where the vector $\vec{x} =(\mu^i) \in M_n $.
Explicitly for the case of two dimensional intrinsic geometry,
the components of the thermodynamic metric tensor are \\
\ba
\label{eq2}
 g_{\mu_1 \mu_1}&=& \frac{\partial^2 F}{\partial \mu_1^2},\nn
 g_{\mu_1 \mu_2}&=& \frac{\partial^2 F}{{\partial \mu_1}{\partial \mu_2}},\nn
 g_{\mu_2 \mu_2}&=& \frac{\partial^2 F}{\partial \mu_2^2}.\nn
\ea Now we can calculate the $\Gamma_{ijk}$, $R_{ijkl}$, $R_{ij}$
and $ R $ for above two dimensional thermodynamic geometry
$(M_2,g)$ and may easily see that the scalar curvature is given by

\ba
\label{eq3}
R&=& -\frac{1}{2} (F_{\mu_1\mu_1}F_{\mu_2 \mu_2}- F_{\mu_1 \mu_2}^2)^{-2}
(F_{\mu_2 \mu_2}F_{\mu_1\mu_1\mu_1}F_{\mu_1 \mu_2 \mu_2}\nn &&+
F_{\mu_1 \mu_2}F_{\mu_1\mu_1\mu_2}F_{\mu_1\mu_2\mu_2}+
F_{\mu_1\mu_1}F_{\mu_1\mu_1\mu_2}F_{\mu_2\mu_2\mu_2}\nn &&-
F_{\mu_1\mu_2}F_{\mu_1\mu_1\mu_1}F_{\mu_2\mu_2\mu_2}-
F_{\mu_1\mu_1}F_{\mu_1\mu_2\mu_2}^2\nn&&
- F_{\mu_2\mu_2}F_{\mu_1\mu_1\mu_2}^2 ).
\ea

Furthermore, the relation between this thermodynamic scalar
curvature and the thermodynamic curvature tensor for such a two
dimensional thermodynamic intrinsic geometry (see for details
\cite{bnt}) is given by \be \label{eq4}
 R=\frac{2}{\Vert g \Vert}R_{\mu_1\mu_2\mu_1\mu_2},
\ee
which is quite usual for any intrinsic Riemannian surface $(M_2(R),g)$.

Note that such Riemannian structures defined by the metric tensor
in a chosen representation are in fact closely related to
classical thermodynamic fluctuation theory \cite {rup2} and
critical phenomena. The probability distribution of thermodynamic
fluctuations in the equilibrium intrinsic space naturally
characterizes the invariant interval of the corresponding
thermodynamic geometry, which in the Gaussian approximation reads
$ \Omega(x)= A \ \ exp [ -\frac {1}{2} g_{ij}(x) dx^i \otimes
dx^j] $, where the pre-factor $A$ is the normalization constant of
the Gaussian distribution. The associated inverse metric may
easily be shown to be the second moment of the fluctuations or the
pair correlation functions and thus may be given as $g^{ij}= < X^i
\vert X^j>$, where  $\lbrace X_i \rbrace $'s are the intensive
thermodynamic variables conjugated to $\lbrace x^i \rbrace$.
Moreover, such Riemannian structures may further be expressed in
terms of any suitable thermodynamic potential obtained by certain
Legendre transform(s) which corresponds to general coordinate
transformations on the equilibrium intrinsic manifold. Our
geometric formulation thus tacitly involves a statistical basis in
terms of a chosen ensemble, although the analysis has only been
considered in the thermodynamic limit of QCD.

We wish to mention that the thermodynamic curvature corresponds to
the nature of the correlation present in the statistical system.
This in particular will imply that the scalar curvature for two
component systems can be thought of as the square of the
correlation length at some given QCD transition temperature $
T_C(\mu_1, \mu_2) $ to be $ R(\mu_1, \mu_2) \sim \xi^2 $, where we
may identify the $ \xi(T_C) $ to be the correlation length  of the
thermodynamic system. In turn, it is not surprising that the geometric
analysis based on a Gaussian approximation to the classical fluctuation
theory precisely yields the correlation volume of QCD. This strongly
suggests that even in the context of chemical reactions in any closed
systems, non-zero scalar curvature might provide useful information
regarding the possible interactions between various components of
the QCD configuration.

Note further that the relation of a non-zero scalar curvature with
an underlying interacting statistical system remains valid even
for higher dimensional intrinsic manifolds and the connection of a
divergent scalar curvature with phase transitions may accordingly
be divulged from the Hessian matrix of the considered QCD free
energy. It is worth to mention that our analysis takes an account
of the scales that are larger than the correlation length and
considers that only few QCD microstates do not dominate the
macroscopic quantities. In particular, we shall focus on the
interpretation that the underlying QCD free energy includes
contributions from a large number of microstates and thus our
description of the geometric QCD thermodynamics. With this general
introduction to the thermodynamic geometry defined as the Hessian
function of the free energy, let us now proceed to investigate the
free energy of hot QCD and its thermo-geometric structures. We now
proceed to compute the QCD free energies in the required form and
thereafter study the thermodynamic geometry of the quark theories
arising from the free energy of hot QCD.

\section{The Free Energy of Hot QCD}

To obtain the free energy as a function temperature and quark
chemical potentials, we consider the quasi-particle model of hot
QCD. We consider two cases in the present paper {\it viz.} the
2-flavor QCD with two distinct chemical potentials ($\mu\sim
(\mu_1,\mu_2)$) and the 3-flavor QCD with two distinct chemical
potentials ($\mu_1$ corresponds to up and down quarks and $\mu_2$
corresponds to strange quark). In order to obtain the expression
for the free energy we consider the following equilibrium
distribution functions for gluons and quarks:

\ba
\label{eq5}
f^{g}_{eq}&=&\frac{1}{(\exp(-\beta\epsilon_p)-1)},\nn
f^{q}_{eq}&=&\bigg\lbrace\frac{1}{(\exp(-\beta\epsilon_p-\beta\mu)+1)}+ \frac{1}{(\exp(-\beta\epsilon_p+\beta\mu)+1)}\bigg\rbrace.\nn
\ea

To obtain the free energy one utilizes the well known
thermodynamic relation,
 \be \label{eq6} F=-\frac{1}{\beta
V} \bigg(\ln(Z_g)+\ln(Z_M)\bigg), \ee
which is nothing but the
negative of the pressure of hot QCD. Here $Z_g$ and $Z_M$ are the
grand canonical partition functions corresponding to the pure
gluonic part and matter sector respectively and have the following
well known form in terms of the gluons and quark-antiquark
distributions functions:

\ba
\label{eq7}
\ln(Z_g)&=& -\sum_{p}\ln(1-exp(1-\exp(-\beta\epsilon_p))),\nn
\ln(Z_M)&=& \sum_{p}\bigg(\ln(1+\exp(-\beta(\epsilon_p-\mu)))\nn&&+\ln(1+\exp(-\beta(\epsilon_p+\mu)))\bigg).
\ea

We now turn our attention to compute the free energy in the small
chemical potential limit $(\beta\vert\mu\vert<<1)$.  We shall determine the
free energy only up to  quartic terms in $\beta\mu$. The
expression thus obtained reads in compact notations as

\be
\label{eq8}
F= T^4\bigg(a_0 +c_1\frac{\mu_1^2}{T^2} + c_2 \frac{\mu_2^2}{T^2} +c^1_4 \frac{\mu_1^4}{T^4}
+c^2_4\frac{\mu_2^4}{T^4}\bigg) +T^4 O(\mu^6/T^6).
\ee

We shall firstly consider the first three terms for our analysis
which we denote as case (A) and the full expression will be
denoted as case (B). We shall study these two cases systematically
and thereby show that the geometric quantities remain intact  at
zero chemical potentials.
On comparison of the free energy displayed in Eq.\ref{eq8} for 2-
and 3-flavors with the temperature dependence of the corresponding
pressure either in lattice QCD \cite{lattice,fkarsch,boyd} or
improved pQCD \cite{htleos,kaj,pqcd,vrn}, one may conclude that
the coefficients of the $\mu_1^2$ and $\mu_2^2$ should be negative
in the corresponding pressure. We shall utilize this to fix the
sign of the various terms in the Taylor expansion of the free
energy as a function of the chemical potentials
($\mu_1$ and $\mu_2$). The various coefficients in Eq.\ref{eq8} for 2 and 3-flavour QCD are as follows:\\

For 2-flavor QCD, \ba \label{eq9}
a_0=-\frac{8\pi^2}{45}\bigg(1+\frac{42}{32}\bigg), c_1=c_2=1/2,
c^1_4\equiv c^2_4=-\frac{1}{4\pi^2}. \ea For 3-flavour case,
the above coefficient reads \ba \label{eq10}
a_0&=&-\frac{8\pi^2}{45}\bigg(1+\frac{63}{32}\bigg), c_1=1,
c_2=1/2, c^1_4= -\frac{1}{2\pi^2}, \nn
 c^2_4&=&-\frac{1}{4\pi^2}.
\ea
Note that, all the above coefficients are dimensionless. Let
us now turn our attention to analyze the free energy of hot QCD
near $T_C$ and trade off the $T_C$ dependence in favour of quark
chemical potentials resulting the following the relation \cite{transition}

\be
\label{eq11}
\frac{T_c(\mu)}{T^0_c}=1+\sum^{2}_{i=1} \tilde {a}_i\frac{\mu_i^2}{(T^0_c)^2}.
\ee

where $ \tilde{ a}_1=\tilde {a}_2=-0.07$ for the 2-flavor case and
$\tilde{a}_1= \tilde{a}_2=-0.114$ for 3-flavor QCD. For the purpose
of the subsequent analysis we choose $T^0_c=0.203$ Gev for $N_f=2$
QCD and $T^0_c=0.197$ Gev in 3-flavor QCD \cite{zantow}. After
substituting for the transition temperature $T_C$, as the function
of quark chemical potentials in the free energy, we thus obtain the
following polynomial form for the free energy: \ba \label{eq12}
F(\mu_1, \mu_2):&=& a_0(b+ a_1 \mu_1^2+ a_2 \mu_2^2)^4+ (b+ a_1
\mu_1^2+ a_2 \mu_2^2)^2\nn && \times (c_1 \mu_1^2+ c_2 \mu_2^2) +
c_4^{(1)} \mu_1^4+ c_4^{(2)} \mu_2^4, \ea where $b=T^0_c$,
$a_1=\tilde{a_1}/b$ and $a_2=\tilde{a_2}/b$. Both $a_1$ and $a_2$
have the dimension of $b^{-1}$(GeV). The free energy will be
measured in units of $GeV^{4}$. It is worth mentining that the
components of the thermodynamic metric tensor are in units of
$GeV^2$, hence the thermodynamic curvature is in units of $GeV^4$,
and the chemical potentials $\mu_i$ are in the units of $GeV$.

The above expression for the free energy is now only a function of the
two quark chemical potentials $(\mu_1,\mu_2)$, since the temperature
dependence has been traded off in favor of the quark chemical potentials.
The validity of the free energy $F(\mu_1, \mu_2)$ as depicted in the
Eq.(\ref{eq12}), and thus all the quantities derived from it depends
upon the condition whether the chemical potentials, $\mu_1$ and $\mu_2$
satisfy: $\vert\mu_i\vert< T^{0}_c$ (i=1,2). We shall show later, the
variation of $\mu_1$ and $\mu_2$ are within the range of validity.

The underlying covariant thermodynamic geometry arising from this
free energy of hot QCD can be determined by employing the general
formalism quoted in the previous section, which we do systematically
in Section 4. Let us first discuss how one incorporates thermal fluctuations
into the free energy and later on we shall study the contributions
to the geometric structure of free energy.

Let us now compute the contribution to the thermodynamic geometry
from the thermal fluctuations. Let us consider case (A) for the
2-flavor QCD first. The free energy defined as the negative
logarithm of the partition function takes the general polynomial
form \be \label{eq13}
 F(\mu_1, \mu_2):= a_0+ c_1 (\frac{\mu_1}{T})^2+ c_2 (\frac{\mu_2}{T})^2. \ee

Since the specific heat as the thermodynamic parameters at
constant volume may be defined as the second partial derivative of
the free energy with respect to the temperature, this enforces
that the associated leading order free energy with the inclusion
of the thermal fluctuations after substituting for $T_C$ as a
function of chemical potentials reads \ba \label{eq14}
 F(\mu_1, \mu_2):&=& a_0+ \frac{c_1 \mu_1^2+ c_2 \mu_2^2}{(b+ a_1 \mu_1^2+ a_2 \mu_2^2)^2}\nn
&&+\frac{1}{2}\ln(\frac{c_1 \mu_1^2+ c_2 \mu_2^2}{(b+ a_1 \mu_1^2+ a_2 \mu_2^2)^2}).
\ea

Let us turn our attention to explore the thermodynamic geometry of
the equilibrium configuration space of the hot QCD, arising from
the free energy expression at the various order perturbative
contributions as well as that of the thermal fluctuations at the
leading order free energy. In order to analyze the geometric
nature of the hot QCD, we shall focus our attention on the
realistic values characterizing certain specific phases of the
QCD. In particular, the examples of interest are the case of two
flavor QCD or that of 3-flavor QCD. As the exact expression for
the components of the metric tensors, associated determinants and
that of the scalar curvatures are rather involved with general
coefficients, thus we shall provide the geometric expressions only
for the realistic parameters of the free energy which describes
the present nature in hot QCD near the $T_C$. For simplicity, we
have not explicitly considered the mass of the quarks but on
general ground, the geometric conclusions may be expected to
retain the same conclusions. Moreover, this is an interesting path
to pursue and, for instance by arguments similar to the one given
above we can conclude that the general results with the inclusion
of the thermal fluctuations agree with the results for arbitrary
perturbatively corrected free energy in the framework of the
theories of the quasi-particles. We shall now proceed to study the
thermodynamic geometry in Case(A) and Case(B) one by one in the 2-
and 3-flavor QCD. We shall henceforth suppress the units of the
thermodynamic metric ($GeV^2$) and thermodynamic curvature
($GeV^4$).

\section{2-flavor QCD}

In this section, we shall present the essential features of
thermodynamic geometries and thereby put them into effect for the
2-flavor hot QCD. In particular, we focus our attention on the
geometric nature of QCD in the neighborhood of $T_C$ with given
quark chemical potentials introduced in the framework of
quasi-particle theories. As stated earlier the thermodynamic
metric in the chemical potential space is given by the Hessian
matrix of the free energy with respect to the intensive variables
which in this case are the two distinct chemical potentials
carried by the quarks.

\subsection{Case(A)}

Considering Eq.\ref{eq12} in the neighborhood of $T_C$ and
substituting for $T_C$ as function of $\mu_1$ and $\mu_2$, one
obtains the following expression: \ba \label{eq15}
 F(\mu_1, \mu_2):&=& a_0(b+ a_1 \mu_1^2+ a_2 \mu_2^2)^4+
(b+ a_1 \mu_1^2+ a_2 \mu_2^2)^2\nn &&\times (c_1 \mu_1^2+ c_2 \mu_2^2).
\ea

To obtain the thermodynamic metric tensor in the chemical
potential space, we employ the formula in Eq. \ref{eq2}, which
leads to the following polynomial expression for the components of
the metric tensor:

\ba
\label{eq16}
g_{\mu_1, \mu_1}&=& 1.12 \mu_2^4+ 5.60 \mu_1^4+ 6.72 \mu_1^2 \mu_2^2
-6.48 \mu_1^4 \mu_2^2\nn &&- 3.89 \mu_1^2 \mu_2^4- 2.21 \mu_1^2- 0.74 \mu_2^2
+ 0.13\nn &&- 3.02 \mu_1^6- 0.43 \mu_2^6,\nn
g_{\mu_1, \mu_2}&=& -1.48 \mu_1 \mu_2+ 4.48 \mu_1^3 \mu_2+ 4.48 \mu_1 \mu_2^3\nn &&
- 2.59 \mu_1^5 \mu_2- 2.59 \mu_1 \mu_2^5- 5.18 \mu_1^3 \mu_2^3,\nn
g_{\mu_2, \mu_2}&=& 1.12 \mu_1^4+ 5.60 \mu_2^4+ 6.72 \mu_1^2 \mu_2^2
-6.48 \mu_1^2 \mu_2^4\nn &&- 3.89 \mu_1^4 \mu_2^2- 2.21 \mu_2^2- 0.74 \mu_1^2
+ 0.13\nn &&- 3.02 \mu_2^6- 0.43 \mu_1^6.
\ea

We thus see that the components of metric functions indeed satisfy
(i) $g_{\mu_1, \mu_1}(\mu_1, \mu_2)= g_{\mu_2, \mu_2}(\mu_2,
\mu_1) $ and (ii) $g_{\mu_1, \mu_2}(\mu_1, \mu_2)= g_{\mu_1,
\mu_2}(\mu_2, \mu_1) $. The thermodynamic geometry predictions
thus obtained indicate that the nature of two point correlation
functions of such 2-flavor QCD must satisfy (i) and (ii). In
particular, this determines the nature of associated quark number
that the susceptibility tensor satisfies: $ \chi_{ij}(\mu_1,
\mu_2) = \chi_{ji}(\mu_2, \mu_1)= \chi $ and $\chi_{ij}(\mu_1,
\mu_2)= \chi_{ij}(\mu_2, \mu_1) $, where $i,j$ are the flavor
indices associated with the 2-flavor QCD configuration, see for a
phenomenological example \cite{hep-ph/0110369v3}. Furthermore, the
determinant of the metric tensor turns out to be polynomial
, and can be given in compact notations as follows

\ba
\label{eq17}
 g(\mu_1,\mu_2)=\sum_{k,l=0 \vert k+l\le 6}^6 a^{A}_{k,l} \mu_1^{2 k}\mu_2^{2l}
 \ea

\begin{figure}[h]
\vspace*{0cm}
\begin{center}
\hspace{-1cm}
\mbox{
\epsfig{file=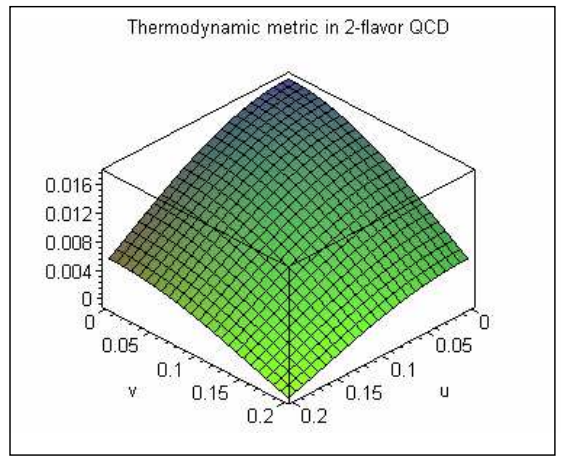,width=6cm,angle=-90}}
\end{center}
\vspace*{0.01cm} \caption{The determinant of the thermodynamic
metric in 2-flavor QCD in the chemical potentials surface. Note
that we measure the chemical potential in GeV.} \vspace*{0.5cm}
\end{figure}

Employing the formula displayed in Eq.\ref{eq4}, we obtain the
thermodynamic curvature which is also a polynomial in $\mu_1$ and $\mu_2$.
It can be written as

\ba
\label{eq18}
 R(\mu_1, \mu_2)= -\frac{4}{g^2}\sum_{k,l=0 \vert k+l\le 7}^7 b^{A}_{k,l} \mu_1^{2k}\mu_2^{2l} .
\ea

\begin{figure}[h]
\vspace*{0cm}
\begin{center}
\hspace{-1cm}
\mbox{
\epsfig{file=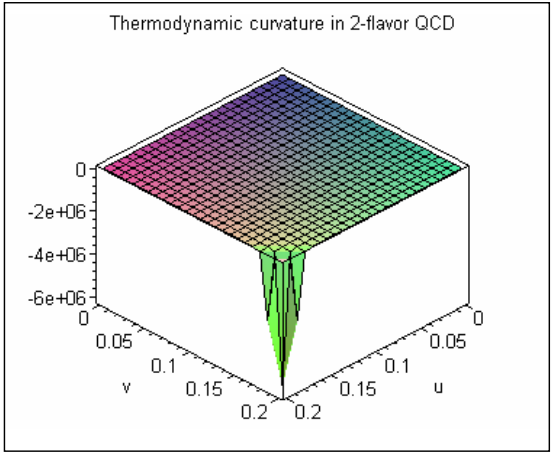,width=6cm,angle=-90}}
\end{center}
\vspace*{0.01cm} \caption{Thermodynamic curvature in 2-flavor QCD
in the chemical potentials surface. Note that we measure the
chemical potential in GeV.} \vspace*{0.01cm}
\end{figure}

The determinants of the thermodynamic metric and curvature in the
chemical potential space are shown in Figs 1 and 2 respectively.
We denote $(\mu_1, \mu_2)\equiv(u,v)$, and henceforth focus our
attention in the physically interesting domain near $T_C$, where
both $u$ and $v$ lie between $0.0$ to $0.2$. In this case, we see
that there are two distinct curves of certain bumps of varying
lengths whose size depends on the domain chosen in the $(u,v)$
space. The significance of these bumps is that they show
thermodynamical interactions in the underlying system. For the
determinant of the metric tensor, when plotted against the
chemical potentials, we find that it is a regular increasing
function of the chemical potentials in the considered physical
domains of hot QCD.

\subsection{Case(B)}
Considering the expansion of the free energy given in Eq.
\ref{eq12}, and employing Eq. \ref{eq2} for non vanishing $c^1_4,
c^2_4 $ defined in Eq. \ref{eq9}, the following expression for the
components of the metric tensor may be obtained:

\ba
\label{eq19}
 g_{\mu_1, \mu_1}&=& 6.72 \mu_1^2 \mu_2^2- 2.52 \mu_1^2 -0.74 \mu_2^2+ 0.13
+5.60 \mu_1^4 \nn && + 1.12 \mu_2^4- 3.02 \mu_1^6
- 0.43 \mu_2^6- 6.48 \mu_1^4 \mu_2^2\nn &&- 3.89 \mu_1^2 \mu_2^4,\nn
g_{\mu_1, \mu_2}&=& -1.48 \mu_1 \mu_2+ 4.48 \mu_1^3 \mu_2+ 4.48 \mu_1 \mu_2^3
- 2.59 \mu_1^5 \mu_2\nn&&- 2.59 \mu_1 \mu_2^5
- 5.18 \mu_1^3 \mu_2^3,\nn
 g_{\mu_2, \mu_2}&=& 6.72 \mu_1^2 \mu_2^2- 2.52 \mu_2^2 -0.74 \mu_1^2+ 0.13
+5.60 \mu_2^4\nn&& + 1.12 \mu_1^4
- 3.02 \mu_2^6- 0.43 \mu_1^6- 6.48 \mu_1^2 \mu_2^4
\nn&&- 3.89 \mu_1^4 \mu_2^2.
\ea

We see further that in this Case(B), the
symmetry($\mu_i\rightarrow-\mu_i$) and exchange symmetry of the
thermodynamic geometry remains conserved as that in the Case(A).
This in fact remains true in general for any symmetric function $
F(\mu_1, \mu_2) $ whose Hessian matrix defines the metric tensor
of the underlying intrinsic Riemannian manifold. The determinant
of the metric tensor turns out to be a polynomial function in the
chemical potentials $ \lbrace \mu_1, \mu_2 \rbrace $, which in
turn is given in the compact notations as

\ba
\label{eq20}
 g(\mu_1, \mu_2)=
\sum_{k,l\vert k+l=0\le 6}^6 a^{B}_{k,l}\ \mu_1^{2l}\mu_2^{2k}
\ea

\begin{figure}[h]
\vspace*{0cm}
\begin{center}
\hspace{-1cm}
\mbox{
\epsfig{file=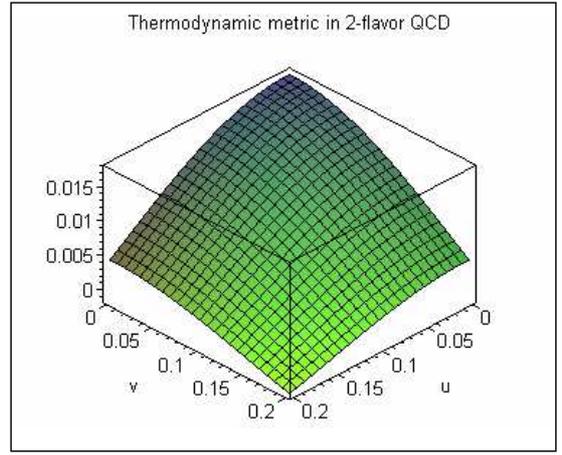,width=6cm,angle=-90}}
\end{center}
\vspace*{0.01cm} \caption{The determinant of the thermodynamic
metric in 2-flavor QCD as a surface plotted against the chemical
potentials. Note that we measure the chemical potential in GeV.}\ba
\label{eq21}
R(\mu_1, \mu_2) =-\frac{4}{g^2} \sum_{k,l=0\vert k+l\le 7}^7 b^{B}_{k,l}\ \mu_1^{2k}\mu_2^{2l}
\ea
\vspace*{0.5cm}
\end{figure}

We see that  the determinant of the metric tensor remains non-zero
in the small chemical potential limit and thus defines a
non-degenerate thermodynamic geometry near $T_C$. Finally, we may
easily obtain the underlying thermodynamic curvature which also having a polynomial
form and can be written as,

\begin{figure}[h]
\vspace*{0cm}
\begin{center}
\hspace{-1cm}
\mbox{
\epsfig{file=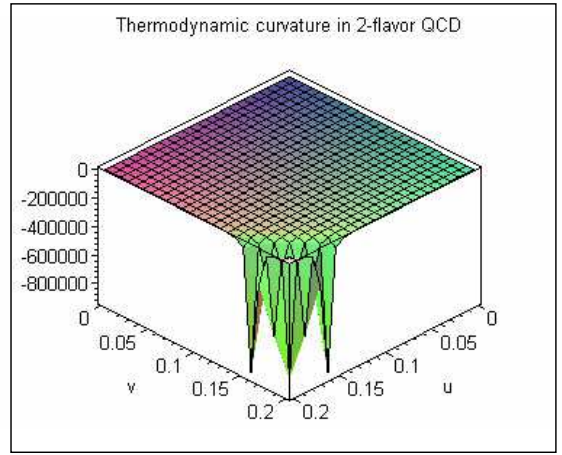,width=6cm,angle=-90}}
\end{center}
\vspace*{0.01cm}
\caption{Thermodynamic curvature in 2-flavor QCD in the chemical potentials surface.
Note that we measure the chemical potentials in GeV.}
\vspace*{0.01cm}
\end{figure}

The behavior of $g$ and $R$ as a function quark chemical potential
potentials in Case (B) for 2-flavor QCD are shown in Figs.3 and 4.
Here the conclusions to be drawn in the range of the chemical
potentials considered in the previous case remain the same, except
for the fact that the curves of the interactions are aligned with
an enhanced magnitude, while the shape of the determinant of the
metric tensor plotted against the chemical potentials remains
(exactly) the same as in the previous case.

We see however that the components of the metric tensor at zero
chemical potentials take the values $ g_{\mu_1,\mu_1}= a^{B}_{0,0}\equiv 0.13=
g_{\mu_2,\mu_2} $ and $ g_{\mu_1,\mu_2}= 0 $, which effectively
describe nothing more than the diagonal and, respectively, off
diagonal quark susceptibility tensors. We find further that the
determinant of the metric tensor reduces to the value of $g= 0.02$
at zero chemical potentials, while the scalar curvature becomes
zero at this point and thus the underlying statistical system at
this point becomes non-interacting. We further see that these
diagonal and off diagonal quark susceptibility tensors, as well as
the determinant of the metric tensor and the scalar curvature at
zero chemical potentials, remain the same under the
$(\frac{\mu}{T})^4$-contributions.

In the case of the 2-flavor QCD free energy, we note that the
degree of the determinant of the thermodynamic geometry in the
chemical potential space remains intact for the Case(B) as that of
the  Case(A). We further see that the associated thermodynamic
curvature never diverges for any values of the chemical
potentials, and tends to a vanishingly small quantity when either
one of the chemical potential corresponds to a large value. Our
predictions can thus easily be generalized to the case when the
free energy may be expressed as a finite polynomial expression of
two chemical potentials. We have further noticed that the
thermodynamic geometry of the free energy preserves all the
symmetries of the free energy. Let us now proceed to study the
effects of thermal fluctuations to the thermodynamic geometry of
case (A) in 2-flavor QCD.

\subsection{Thermal Fluctuations}

We will now discuss the thermodynamic geometry of hot QCD with the
inclusion of thermal fluctuations about an equilibrium. As it is
well known, any thermodynamical system considered as a canonical
ensemble has logarithmic and polynomial contributions to the free
energy \cite{SM}. These considerations apply as well to
quasi-particles models of QCD, and the specific forms of the
logarithmic and polynomial contributions may easily be calculated
for a wide class of particle theories. In order to the analyze
essential geometric nature of hot QCD under thermal contributions,
we shall consider the case of the quadratic corrected free energy.
Further it is straightforward to see that the case of quartic
corrected free energy in this framework leaves the qualitative
nature of results untouched. This approximation is valid only in
the regime where thermal fluctuations are much larger than quantum
fluctuations.

With the above consideration in the leading order quadratic
approximation, the corresponding free energy of two flavor QCD
reads

\ba
\label{eq22}
F(\mu_1, \mu_2)&=& a_0+ \frac{c_1 \mu_1^2+ c_2 \mu_2^2}{(b+ a_1 \mu_1^2+ a_2 \mu_2^2)^2} \nn &&
+ \frac{1}{2} \ln(\frac{c_1 \mu_1^2+ c_2 \mu_2^2}{(b+ a_1 \mu_1^2+ a_2 \mu_2^2)^2}).
\ea

Employing the formula to determine the metric tensor and
thermodynamic curvature displayed in Eqs.\ref{eq2}, the components
are the metric tensor in the scale of $10^{10} $ come out to be

\ba
\label{eq23}
 g_{\mu_1, \mu_1}&=& \lbrace -0.17 (\mu_1^2- \mu_2^2)
 -1.33 (\mu_2^{10}-\mu_1^{10})\nn && + 9.96 \mu_2^{8}
-36.26 \mu_1^{8} -3.55 \mu_2^{4}- 5.82\mu_1^{4}\nn &&
+2.67 (\mu_2^{6}\mu_1^{4}- \mu_1^{6}\mu_2^{4})
-102.80 (\mu_2^{2}\mu_1^{4}- 0.5 \mu_1^{2}\mu_2^{4})\nn &&
-78.93 \mu_1^{4}\mu_2^{4}- 36.27 \mu_1^{8}
-51.40 \mu_1^{6}- 5.82 \mu_1^{4}\nn &&- 1.33 \mu_1^{10}
-6.38 \mu_2^{6}\mu_1^{2} - 98.86 \mu_2^{2}\mu_1^{6}\nn &&
-4.00 (\mu_2^{2}\mu_1^{8}- \mu_1^{2}\mu_2^{8})
-9.37 \mu_1^{2}\mu_2^{2} \rbrace\nn&&\times
(-203+340\mu_1^2+340\mu_2^2)^{-4}(\mu_1^2+ \mu_2^2)^{-2},
\nn
g_{\mu_1, \mu_2}&=& 2 \mu_1 \mu_2 \lbrace 1.13 (\mu_1^2+ \mu_2^2)
+23.12 (\mu_2^6+\mu_1^6)\nn &&+ 1.33 (\mu_1^8+ \mu_2^8)+
51.40(0.50 \mu_1^4+ \mu_2^4)\nn &&+ 69.36 (\mu_2^2\mu_1^4+ \mu_1^2\mu_2^4)
8.01(\mu_1^4+ \mu_2^4)\nn &&+ 5.34 (\mu_1^2\mu_2^6+ \mu_1^6\mu_2^2)
+ 51.40 \mu_1^2\mu_2^2- 0.17 \rbrace \nn &&\times
(-203+340\mu_1^2+340\mu_2^2)^{-4}(\mu_1^2+ \mu_2^2)^{-2},
\nn
g_{\mu_2,\mu_2}&=& \lbrace -0.17 (\mu_2^2- \mu_1^2)
-1.33 (\mu_1^{10}-\mu_2^{10})\nn && + 9.96 \mu_1^{8}
-36.26 \mu_1^{8} -3.55 \mu_2^{4}- 5.82\mu_1^{4}\nn &&
+2.67 (\mu_1^{6}\mu_2^{4}- \mu_2^{6}\mu_1^{4})
-102.80 (\mu_1^{2}\mu_2^{4}\nn &&- 0.5 \mu_2^{2}\mu_1^{4})
-78.93 \mu_2^{4}\mu_1^{4}- 36.27 \mu_2^{8}\nn &&
-51.40 \mu_2^{6}- 5.82 \mu_2^{4}- 1.33 \mu_2^{10}
-6.38 \mu_1^{6}\mu_2^{2}\nn &&- 98.86 \mu_1^{2}\mu_2^{6}
-4.00 (\mu_1^{2}\mu_2^{8}- \mu_2^{2}\mu_1^{8})\nn &&
-9.37 \mu_2^{2}\mu_1^{2} \rbrace \times
(\mu_1^2+ \mu_2^2)^{-2}\nn &&
(-203+340\mu_1^2+340\mu_2^2)^{-4}.
\ea

It is thus evident that the symmetry of the metric tensor remains
preserved under thermal fluctuations. In this case, after
incorporating the Log contributions to the free energy in the case
of 2-flavor QCD, we obtain the following positive definite
expressions for the determinant of the thermodynamic metric:
 \ba
\label{eq26}
g(\mu_1,\mu_2)= 4 \frac{\sum_{k,l=0\vert k+l\le 7}^7 a^{C}_{l,k}\  \mu_1^{2k} \mu_2^{2l}}
{(\mu_1^2+\mu_2^2))^{2}(0.20-0.34(\mu_1^2-\mu_2^2))^{7}}
 \ea
\begin{figure}[h]
\vspace*{0cm}
\begin{center}
\hspace{-1cm}
\mbox{
\epsfig{file=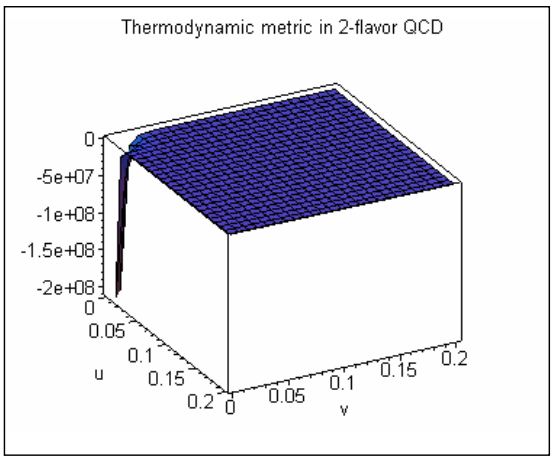,width=6cm,angle=-90}}
\end{center}
\vspace*{0.01cm} \caption{The determinant of the metric tensor in
2-flavor QCD under thermal corrections in the chemical potentials
surface. Note that we measure the chemical potential in GeV.}
\vspace*{0.5cm}
\end{figure}

We see in this case again that the thermodynamic curvature turns
out in the form of the ratio of two polynomial expressions, which
may be written as

\ba
\label{eq27}
 R(\mu_1, \mu_2)&=& -\frac{4}{g^2} ( 0.20 -0.34 (\mu_1^2 +\mu_2^2)\nn&& \times
\sum_{k,l=0 \vert k+l\leq 12}^{12} b^{C}_{k,l}\ \mu_1^{2 k}\mu_2^{2l}, \ea

where the coefficients $ \lbrace a_{i,j} \rbrace $ of
the polynomial expression appearing in the numerator may easily be
tracked from the corresponding points on the figure 6. In this
case, we should note that the determinant of the metric tensor
plotted against the chemical potentials becomes a smoother
function in comparison to that of the case without logarithmic
correction. However, we observe that it reaches a magnitude of
$10^9$ near the $(u,v)=(0.2,0.2)$. Further, we find that the
domain of the interaction has shifted towards the origin. In
particular, we see that such interactions are only present in the
chemical potentials range $0.1 < \mu_i < 0.2$.
\begin{figure}[h]
\vspace*{0cm}
\begin{center}
\hspace{-1cm}
\mbox{
\epsfig{file=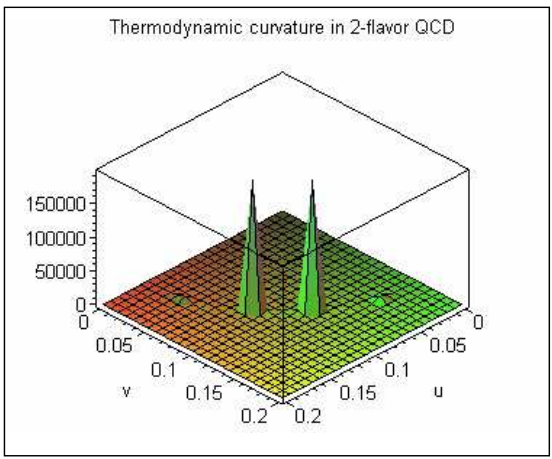,width=6cm,angle=-90}}
\end{center}
\vspace*{0.01cm} \caption{Thermodynamic curvature in 2-flavor QCD
under thermal corrections in the chemical potentials surface. Note
that we measure the chemical potential in GeV.} \vspace*{0.01cm}
\end{figure}

Finally, we wish to point out a general feature of the above
geometric considerations, i.e. that any polynomial free energy after
including the logarithmic contributions is going to generate
certain bump(s) in the intrinsic Riemannian space of chemical
potentials for a given non-zero $T_C$ around the equilibrium QCD
thermodynamic configuration. Furthermore, following our general
procedure it may explicitly be verified, in the same way as in the
cases of quadratic or quartic contributions to the free energy,
that the logarithmic contributions to the free energy do not spoil
the symmetry of the determinant and that of the associated scalar
curvature of the intrinsic space spanned by the quasi-particle
chemical potentials. Such symmetries are indeed the artifact of
the choice of the parameters appearing in the corresponding free
energy such that they correspond to the symmetric free energy in
the chemical potentials. Due to the inherent symmetries in the
definition of the quasi-particle free energies, it is possible in
principle that we can give up manifestly invariant expressions for
the determinants and the scalar curvatures. In particular, we find
that for all quasi-particle free energy $ F(\mu_1, \mu_2):=
\sum_{i,j} a_{i,j}\mu_1^i \mu_2^j $ with $ a_{i,j}= a_{j,i} $, it
turns out that the determinant and associated scalar curvature of
the intrinsic Riemannian manifold as a function of the quark
chemical potentials are symmetric functions $ g(\mu_1, \mu_2)=
g(\mu_2, \mu_1) $ and $ R(\mu_1, \mu_2)= R(\mu_2, \mu_1) $ of the
chemical potentials.

\section{3-flavor QCD}

As in the previous section, we shall analyze the free energy as a
function of quark chemical potentials with or without thermal
corrections, and thereby systematically explore the covariant
metric tensor, determinant and scalar curvature of the
thermodynamic geometry of 3-flavor QCD.

\subsection{Case(A)}

In what follows next in this subsection, we present the geometric
analysis for the realistic quasi-particle model of QCD near the
$T_{C}$. The thermodynamic geometry of interest may thus easily be
obtained, as before, from the Hessian matrix of the free energy
with respect to the quark chemical potentials.
In this case the components of the metric tensor reduce to\\
\ba
\label{eq28}
 g_{\mu_1, \mu_1}&=&
-7.59 \mu_1^2- 2.30 \mu_2^2+ 0.26+ 33.76 \mu_1^4\nn &&
+6.08 \mu_2^4- 32.50 \mu_1^6+ 38.51 \mu_1^2 \mu_2^2\nn &&
-69.65 \mu_1^4 \mu_2^2 - 41.79 \mu_1^2 \mu_2^4- 4.64 \mu_2^6, \nn
 g_{\mu_1, \mu_2}&=&
-4.60 \mu_1 \mu_2+ 25.67 \mu_1^3 \mu_2+ 24.33  \mu_1 \mu_2^3 \nn &&
-27.86  \mu_1^5 \mu_2- 55.72  \mu_1^3 \mu_2^3- 27.86  \mu_1 \mu_2^5, \nn
 g_{\mu_2, \mu_2}&=&
-6.22 \mu_1^2- 2.30 \mu_2^2+ 0.22+ 6.41 \mu_1^4\nn &&
+28.75 \mu_2^4 - 4.64 \mu_1^6+ 36.50 \mu_1^2 \mu_2^2 \nn &&
-41.79 \mu_1^4 \mu_2^2 - 69.65 \mu_1^2 \mu_2^4- 32.50 \mu_2^6.
\ea

We see further that the components of the metric tensor at zero
chemical potentials take the values $ g_{\mu_1,\mu_1}= 0.26$,
$g_{\mu_2,\mu_2}=0.22 $ and $ g_{\mu_1,\mu_2}= 0 $, which in
effect respectively describe the diagonal and off diagonal quark
susceptibility tensors. We further find that the determinant of
the metric tensor reduces to the value of $g= 0.06$ at zero
chemical potentials, while the scalar curvature vanishes at this
point and thus the underlying statistical system approaches a
non-interacting system. On the other hand, the determinant of the
metric for this 3 flavor case is as follows:

\ba
\label{eq29}
 g(\mu_1, \mu_2)=
\sum_{k,l=0\vert k+l\le 6}^6 \tilde{a}^{A}_{k,l}\ \mu_1^{2l}\mu_2^{2k}
\ea

\begin{figure}[h]
\vspace*{0cm}
\begin{center}
\hspace{-1cm}
\mbox{
\epsfig{file=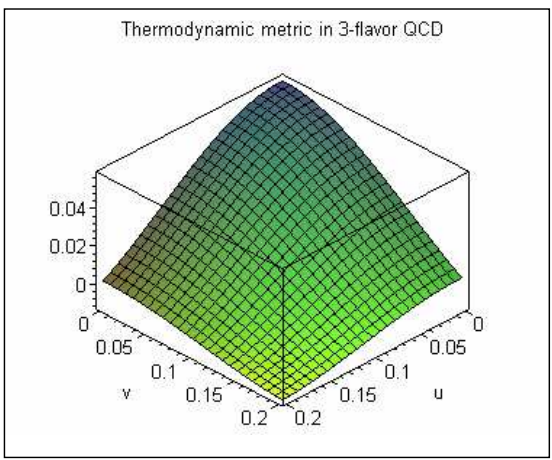,width=6cm,angle=-90}}
\end{center}
\vspace*{0.01cm} \caption{The determinant of the thermodynamic
metric in 3-flavor QCD in the chemical potentials surface. Note
that we measure the chemical potential in GeV.} \vspace*{0.5cm}
\end{figure}

In this case  we see that $g(\mu_1,\mu_2)\neq g(\mu_2,\mu_1)$,
which follows from the fact that the 3-flavor QCD free energy
(displayed in Eq.\ref{eq12}) is not symmetric under the exchange
in the chemical potentials. The same remains true for the
associated curvature, which may be seen from

\ba
\label{eq30}
R(\mu_1, \mu_2) =-\frac{4}{g^2} \sum_{k,l=0\vert k+l\le 7}^7 \tilde{b}^{A}_{k,l}\ \mu_1^{2k}\mu_2^{2l}
\ea

\begin{figure}[h]
\vspace*{0cm}
\begin{center}
\hspace{-1cm}
\mbox{
\epsfig{file=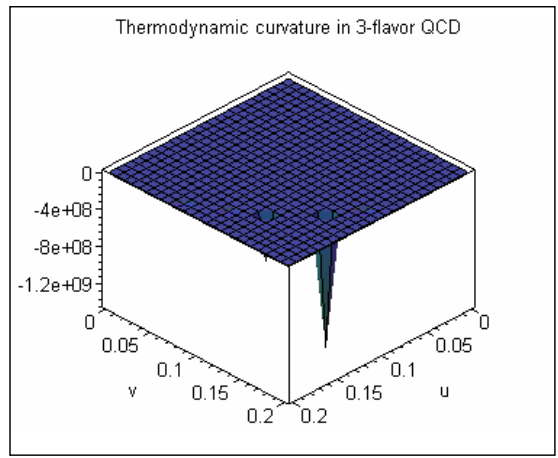,width=6cm,angle=-90}}
\end{center}
\vspace*{0.01cm} \caption{Thermodynamic curvature in 3-flavor QCD
in the chemical potentials surface. Note that we measure the
chemical potential in GeV.} \vspace*{0.01cm}
\end{figure}

From Figs.7 and 8, we see that the determinant of the metric
tensor plotted against the chemical potentials shows two lines of
minima and the maximum lying in between the minima occurs when
either one of the chemical potentials reaches a non-zero constant
value, while the other is fixed at the origin. In this case we
further see that there are two distinct bumps of varying heights
instead of the curves of interactions. The height of the bumps
depends on the domain chosen in the $(u,v)$ space. The
significance of these bumps is that they show thermodynamical
interactions in the system of the 3-flavor QCD.

\subsection{Case(B)}

In this subsection, we shall investigate the thermodynamic
geometry corresponding to the free energy given by Eq.\ref{eq8}.
It turns out here that the components of the metric tensor thus
obtained are as follows:

\ba
\label{eq31}
 g_{\mu_1, \mu_1}&=&
-38.51 \mu_1^2 \mu_2^2+ 6.08 \mu_2^4- 69.65 \mu_1^4 \mu_2^2 \nn&&
-8.20 \mu_1^2- 2.30 \mu_2^2+ 0.61+ 33.76 \mu_1^4\nn&&
- 41.79 \mu_1^2\mu_2^4-32.50 \mu_1^6- 4.64 \mu_2^6,\nn
g_{\mu_1, \mu_2}&=&
-4.60 \mu_1 \mu_2+ 25.67 \mu_1^3 \mu_1+ 24.33 \mu_1 \mu_2^3\nn&&
- 27.86 \mu_1^5 \mu_2
-55.72 \mu_1^3 \mu_2^3- 27.86 \mu_1 \mu_1^5,\nn
 g_{\mu_2, \mu_2}&=&
-36.51 \mu_1^2 \mu_2^2+ 28.75 \mu_2^4- 41.79 \mu_1^4 \mu_2^2\nn&&
-2.30 \mu_1^2- 6.53 \mu_2^2+ 0.22+ 3.42 \mu_1^4\nn&&
- 69.65 \mu_1^2\mu_2^4-4.64 \mu_1^6- 32.50 \mu_2^6.
\ea

In this case the conclusion to be drawn in the range of the
chemical potentials considered in the previous case remains the
same, except the fact that the number of bumps is increased, which
thus indicates stronger interactions in  the internal space of the
quark chemical potentials, while the shape of the determinant of
the metric tensor plotted against the chemical potentials is a bit
flatter near the $(u,v)=(0.2,0.2)$ than the previous case. The
determinant of the metric tensor turns out to be polynomial

\ba
\label{eq32}
 g(\mu_1, \mu_2)=
\sum_{k,l=0 \vert k+l\le 6}^6 \tilde{a}^{B}_{k,l}\ \mu_1^{2l}\mu_2^{2k}
\ea
\begin{figure}[h]
\vspace*{0cm}
\begin{center}
\hspace{-1cm}
\mbox{
\epsfig{file=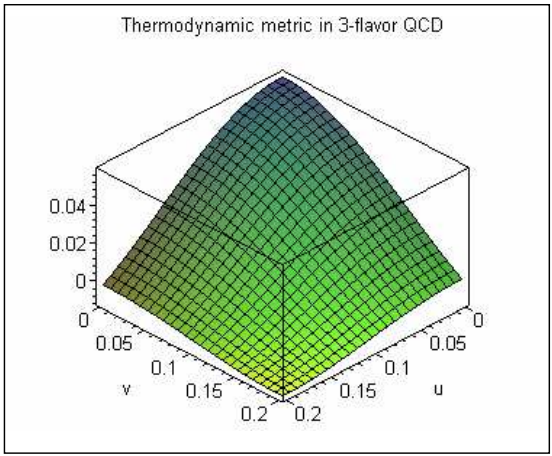,width=6cm,angle=-90}}
\end{center}
\vspace*{0.01cm} \caption{The determinant of the metric tensor in
3-flavor QCD as a function of the chemical potentials. Note that
we measure the chemical potential in GeV.} \vspace*{0.5cm}
\end{figure}

Note further in this case, we again see that $g(\mu_1,\mu_2)\neq
g(\mu_2,\mu_1)$, which follows directly from the fact that the
underlying free energy with $(\frac{\mu}{T})^4$-contributions
(displayed in Eq.\ref{eq8}) remains non-symmetric under the
exchange of the quarks chemical potentials. Thus the same remains
true for the associated curvature as the polynomial ratio of the
Riemann tensor and the determinant of the metric tensor, which
turns out to be given by

\ba
\label{eq33}
R(\mu_1, \mu_2) =-\frac{4}{g^2} \sum_{k,l=0\vert k+l\le 7}^{7} \tilde{b}^{B}_{k,l}\ \mu_1^{2k}\mu_2^{2l}
\ea

\begin{figure}[h]
\vspace*{0cm}
\begin{center}
\hspace{-1cm}
\mbox{
\epsfig{file=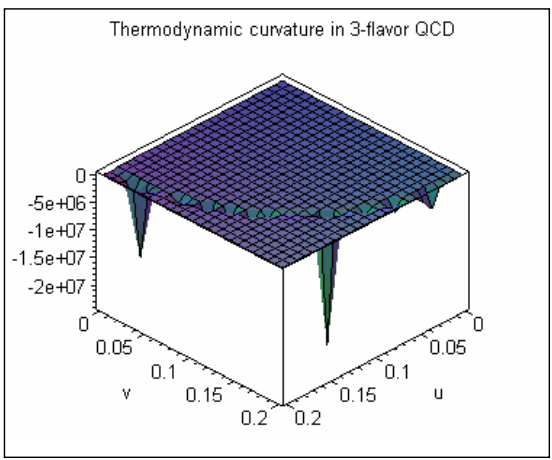,width=6cm,angle=-90}}
\end{center}
\vspace*{0.01cm} \caption{Thermodynamic curvature in 2-flavor QCD
in the chemical potentials surface in case B. Note that we measure
the chemical potential in GeV.} \vspace*{0.01cm}
\end{figure}

In these cases when the free energy may be treated as a function
of the chemical potentials, we see that the thermodynamic space
spanned by the chemical potentials remains completely regular
except at those points for which the determinant of the metric
tensor vanishes on the intrinsic manifold. This intrinsic space of
the chemical potentials thus turns out to be a well-defined,
stable configuration for all $\lbrace \mu_1, \mu_2 \rbrace$ such
that the the determinant of the associated metric tensor remains
positive definite in the domain of interest. We see in this case
that the underlying diagonal quark susceptibility tensor does not
remain the same at zero chemical potentials, and in particular, we
find that $ g_{\mu_1,\mu_1}= 0.61, g_{\mu_2,\mu_2}=0.22 $ and $
g_{\mu_1,\mu_2}= 0 $ while the determinants of the metric tensor
and the scalar curvature remain the same under the
$(\frac{\mu}{T})^4$-contributions at zero chemical potentials.

\subsection{Thermal Fluctuations}

We will now discuss the thermodynamic properties of 3-flavor QCD
under thermal fluctuations, treating the system as a well-defined
ensemble. We allow for small thermal fluctuations in the system
considered, and study the thermodynamic geometry of 3-flavor QCD
in the lines with our treatment of the usual two flavor QCD
described earlier in the previous section. For simplicity we shall
work in the leading order approximation with the corresponding
free energy as that of two flavor QCD, with the understanding that
the parameters appearing in the free energy are given as that of
the quadratic case of 3-flavor QCD. A straightforward computation
yields that the components of the metric tensor in the scale of
$10^{10} $ read

\ba
\label{eq34}
 g_{\mu_1, \mu_1}&=& -2 (-197+ 578 \mu_1^2+ 578 \mu_2^2)^{-4}(2 \mu_1^2+ \mu_2^2)^{-2}\nn &&
\lbrace 7.50 \mu_2^6 + 3.80 \mu_2^8+ 0.30 \mu_1^2-0.15 \mu_2^2\nn &&
- 2.56 \mu_2^4+ 182.75 \mu_1^4 \mu_2^4-28.69 \mu_1^2 \mu_2^4\nn &&
+ 78.13 \mu_1^6 \mu_2^4- 22.32 \mu_1^{10}-416.12 \mu_1^8\nn &&
- 343.63 \mu_1^6 - 11.16 \mu_1^8 \mu_2^2 -91.30 \mu_1^6 \mu_2^2\nn &&
- 19.06 \mu_1^2 \mu_2^2- 274.74 \mu_1^4 \mu_2^2-20.83 \mu_1^4\nn &&
+ 62.18 \mu_1^2 \mu_2^6+ 100.45 \mu_1^4 \mu_2^6
+ 33.48 \mu_1^2 \mu_2^8 \rbrace, \nn
 g_{\mu_1, \mu_1}&=& 4 \mu_1 \mu_2 (-197+ 578 \mu_1^2+ 578 \mu_2^2)^{-4}(2 \mu_1^2+ \mu_2^2)^{-2}\nn &&
\lbrace 7.61 \mu_2^6 + 1.77 \mu_1^2+ 1.77 \mu_2^2+27.68 \mu_2^4\nn &&
+ 78.13 \mu_1^4 \mu_2^4+ 107.83 \mu_1^2 \mu_2^4+33.48 \mu_1^8\nn &&
+ 385.68 \mu_1^6+ 89.29 \mu_1^6 \mu_2^2+126.27 \mu_1^2 \mu_2^2\nn &&
+ 385.68 \mu_1^4 \mu_2^2+ 143.05 \mu_1^4+22.32 \mu_1^2 \mu_2^6\nn &&
- 0.15 \rbrace,\nn
 g_{\mu_2, \mu_2}&=& (-197+ 578 \mu_1^2+ 578 \mu_2^2)^{-4}(2 \mu_1^2+ \mu_2^2)^{-2}\nn &&
\lbrace 80.72 \mu_2^6+ 107.83 \mu_2^8+ 0.30 \mu_2^2-0.15 \mu_2^2\nn &&
- 11.16 \mu_2^{10}+ 6.53 \mu_2^4+1911.89 \mu_1^4 \mu_2^4\nn &&
+ 368.99 \mu_1^2 \mu_2^4+ 66.97 \mu_1^4 \mu_2^6-69.97 \mu_1^{10}\nn &&
- 340.04 \mu_1^8+ 75.53 \mu_1^6-100.54 \mu_1^8 \mu_2^2\nn &&
+ 1278.78 \mu_1^6 \mu_2^2+ 17.29 \mu_1^2 \mu_2^2+437.31 \mu_1^4 \mu_2^2\nn &&
+ 15.52 \mu_1^4+ 801.80 \mu_1^2 \mu_2^6+178.58 \mu_1^4 \mu_2^6\nn &&
+ 89.29 \mu_1^8 \mu_2^2\rbrace.
\ea

The determinant of the metric tensor turns out to be given by the
following polynomial expression:

\ba
\label{eq35}
 g(\mu_1,\mu_2)&=&
 \bigg((0.197-0.578(\mu_1^2+\mu_2^2))^{7} (\mu_1^2+0.50\mu_2^{2})^{2} \bigg)\nn&& \times
\sum_{k,l=0 \vert k+l\leq 12}^{12} \tilde{b}^{C}_{k,l}\ \mu_1^{2 k}\mu_2^{2l}, \ea

\begin{figure}[h]
\vspace*{0cm}
\begin{center}
\hspace{-1cm}
\mbox{
\epsfig{file=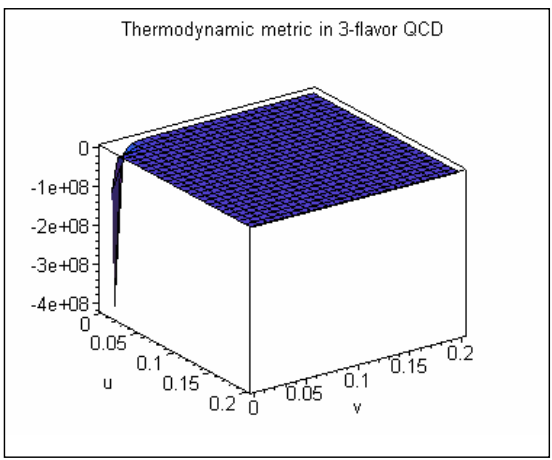,width=6cm,angle=-90}}
\end{center}
\vspace*{0.01cm} \caption{The determinant of the metric tensor in
3-flavor QCD in the chemical potentials surface when the thermal
contributions are included. Note that we measure the chemical
potential in GeV.} \vspace*{0.5cm}
\end{figure}

In this case, we see that the determinant of the metric tensor
plotted against the chemical potentials acquires two finite size
bumps which were absent in the case without thermal corrections.
Again the height of the bumps depends on the location being
chosen. Furthermore, we find that the domain of the interaction
has shifted towards the origin, as in the previously treated case
of 2-flavor QCD. In particular, we see that there are two bumps of
interactions present near the value of either chemical potentials
$ \mu_i =0.1$.  However, we observe that the strength of the
interaction depends on the exact location chosen in the internal
space, for example it may become as high as $10^4$ or as small as
$10^2$ near the $(u,v)=(0.1,0.1)$.

\begin{figure}[h]
\vspace*{0cm}
\begin{center}
\hspace{-1cm}
\mbox{
\epsfig{file=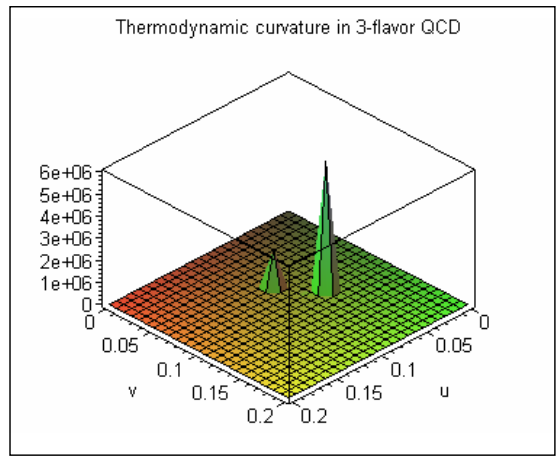,width=6cm,angle=-90}}
\end{center}
\vspace*{0.01cm} \caption{Thermodynamic curvature in 3-flavor QCD
in the chemical potentials surface when the thermal contributions
are included. Note that we measure the chemical potential in GeV.}
\vspace*{0.01cm}
\end{figure}

In this case again it turns out that the thermodynamic curvature
may be written in the same form as the ratio of two polynomial
expressions, and in particular we find that this curvature is
given by
 \ba \label{eq36} R(\mu_1, \mu_2)&=&
-\frac{4}{g^2}(0.197-0.578(\mu_1^2+\mu_2^2))\nn&& \sum_{\lbrace
k,l=0 \vert k+l\leq 12\rbrace}^{12} \tilde{b}^C_{k,l}\ \mu_1^{2l} \mu_2^{2k}, \ea

where the associated coefficients $ \lbrace \tilde{b}_{k,l} \rbrace $ of
the Riemann tensor appearing in the numerator of the scalar
curvature may easily be read out from the corresponding Fig. 12.
Thus, in this section once again one can always find from the
general considerations of the intrinsic Riemannian geometry as in
the case of 2-flavor QCD that for any given polynomial free energy
after including the logarithmic contributions there exist certain
bump(s) in the associated intrinsic Reimannian space.

\section{Discussion and Conclusion}

In this paper we have introduced an intrinsic geometric notion to
QCD thermodynamics, and thereby it has been applied to study the
behavior of $2$-flavor and $3$-flavor QCD in several simple
thermodynamic considerations. The free energy of
the $2$-flavor, as well as that of the 3-flavor QCD, with or
without logarithmic contributions under thermal fluctuations,
indicate an intriguing relationships between the geometrical
concepts of QCD thermodynamics, and in particular, the scalar
curvature of the underlying metric tensor to the correlation
volume. Thus the physical concepts of a phase transition(s), if
any, as well as the geometric apprehension of the quasi-particles,
non-ideal interactions are divulged. For $2$-flavor QCD as a
thermodynamic system with two distinct quark chemical potentials,
the associated quarks susceptibility tensors may consequently be
defined as the components of the thermodynamic metric tensor
arising from the Hessian of the hot QCD free energy and thus
represent the scalar curvature of the underlying geometry from the
viewpoint of the statistical theory of quark states. We have
explicitly determined such notions for a few simple cases, in
particular for the $2$-flavor and the $3$-flavor theories, and it
turns out that they indicate certain physical behavior in the
lieu of \textit{``macroscopic/microscopic duality''} relation(s),
which are indeed in close connection with the \textit{``AdS/CFT
correspondence''} via an associated fermion number density of the
underlying effective field theory. In fact, this study provides
convincing examples that the intrinsic geometric approach analyzes
the thermodynamics of hot QCD systems from the perspective of
microscopic theory and thus can be applied to actually divulge
important physical and chemical behavior of the quasi-particles.

\subsection{Thermodynamic Parameters}
We have shown that the covariant intrinsic geometric notion about
an equilibrium thermodynamics configuration, when applied to the
2-flavor and 3-flavor quasi-particle QCD models considered as
thermodynamic systems, provided that they are an integral part of
the environment with which they are treated to be in equilibrium,
in the framework of an underlying ensemble theory, indicates in
both cases a well-defined, curved, regular intrinsic Riemannian
manifold. Hence, the study of the equilibrium thermodynamic
geometry as an intrinsic manifold arising from quasi-particles and
the associated scalar curvature have led us to interesting
insights into phase transitions and critical phenomena, if any,
for such quasi-particles. However, it may be mentioned that a
thermodynamic fluctuation approach is not viable for a most
general non-perturbatively completed QCD in the framework of
non-equilibrium statistical mechanics with an infinite extensive
environment, which we would like to investigate in the near
future. The basic idea involved here is that the positivity of the
thermodynamic metric tensor cannot be ensured for the full range
of all parameters $\lbrace \mu_i, T, V, S \rbrace$, which is
necessary for a local minimum of the energy for any given
equilibrium configuration. Thus, it is plausible to investigate
such notions in a restricted fashion where fluctuations of one or
more thermodynamic parameter is either frozen out or assumed to be
negligible, see for example \cite{Tawfik} in the context of QCD.

\subsection{Hot QCD near $T_{C}$}
In this paper, we have introduced such geometries to the
thermodynamics of hot QCD, and focused our attention on the
intrinsic Riemannian geometry in the neighborhood of
$T_C(\mu_1,\mu_2)$ of the $2$-flavor, as well as of the $3$-flavor
hot QCD. The geometry thus defined on a two dimensional manifold
is solely spanned by the chemical potentials of quasi-particles
and in turn determines the two point correlation functions and an
associated correlation volume, near the considered transition
temperature of the underlying microscopic theory. Note further
that all those QCD-diagrams which just have one gluon exchange do
not appear in the thermodynamic geometry, which is due to the fact
that such diagrams vanish under color neutrality transformations.
Whereas, the corresponding diagrams with two gluon exchange remain
non-zero, because the fermion loops are even functions of the
chemical potentials and thus contribute to the off-diagonal
susceptibility tensors only at zero chemical potentials. We thus
find that components of the off-diagonal susceptibility tensor
macroscopically correspond to those of the thermodynamic metric
tensor $\lbrace g_{ij} \vert i \neq j \rbrace $, which may further
be envisaged in terms of the fermion number density as
$g_{ij}\equiv \chi_{ij}= \frac{\partial \mathcal N_i}{\partial
\mu_j} $. Our geometric study thus describes the mixing
between different flavors induced by the resummation of quark
loops along the soft, internal, gluon lines to certain orders of
given QCD diagrams.

\subsubsection{2-flavor Hot QCD}
It is worth to point out in the case of $2$-flavor QCD that the
components of the diagonal quark susceptibility tensor and those
of the off-diagonal quark susceptibility tensor may easily be read
off just from the components of the associated thermodynamic
metric tensors at zero quark chemical potentials. In particular,
we see that the diagonal quark susceptibility tensor components $
g_{\mu_i,\mu_i} $ take identical values, while those of the
off-diagonal quark susceptibility tensor identically vanish. It is
straightforward to see that the determinant of the metric tensor
at zero chemical potentials takes a positive definite value and
thus defines a well-defined thermodynamic geometry, while the
associated scalar curvature at this value of the chemical
potentials turns out to be zero, and thus the underlying
statistical system at this point becomes a non-interacting system.
It turns out that both the diagonal and off-diagonal quark
susceptibility tensors, the determinant and the scalar curvature
of such associated metric tensor at zero chemical potentials
remain intact under the $(\frac{\mu}{T})^4$-contributions.

It is worth to note that the geometrical results of QCD, when
plotted in a physically interesting, near $T_C^0$, range against
the chemical potentials are in turn very illuminating. In
particular, for the case of $2$-flavor QCD we find that there are
two distinct curves of bumps of varying strength of
thermodynamical interactions, whose size thus depends on the
chosen domain in the chemical potential space of the system.
Furthermore, the determinant of the metric tensor, when plotted
against the chemical potentials in the considered domains of hot
QCD, has been observed to be a regular increasing function of the
chemical potentials. Interestingly, the conclusions to be drawn in
this range of the chemical potentials remain the same under the
$(\frac{\mu}{T})^4$-contributions and it turns out that the shape
of the determinant of the metric tensor against the chemical
potentials remains exactly the same as that without such
contributions. However, the curves of the interactions present get
aligned with an enhanced magnitude compared to that without
$(\frac{\mu}{T})^4$-contributions for $2$-flavor hot QCD. It is
further important to note that the determinant of the metric
tensor when plotted against the chemical potentials becomes a
smoother function under the thermal contributions. Under the
latter the domain of the interactions is observed to be shifted
towards the origin, and in particular they are dominant in the
range when both chemical potentials lie within $0.1 < \mu_i < 0.2$.

\subsubsection{3-flavor Hot QCD}
In this case, we observe the same for the case of 3-flavor QCD, i.e.
that at zero chemical potentials the components of the metric
tensor $ g_{\mu_i,\mu_i} $  reduce to identical values and in fact
vanish corresponding to the case of $ g_{\mu_1,\mu_2}$, which thus
respectively describe the diagonal and the off-diagonal quark
susceptibility tensors. Moreover, we find that the determinant
of the metric tensor acquires a non-zero positive definite value,
while the associate scalar curvature turns out again to be zero,
and thus the underlying QCD system turns out to be a
non-interacting statistical system at zero quarks chemical
potentials. It is important to note in this case that the diagonal
quark susceptibility tensor components at zero chemical potentials
do not find the same values under the
$(\frac{\mu}{T})^4$-contributions but rather we observe $
g_{\mu_1,\mu_1}= 0.61, g_{\mu_2,\mu_2}=0.22 $ and $
g_{\mu_1,\mu_2}= 0 $. This further implies that the associated
quark susceptibility matrix is non-trivial, and thus the different
quarks acquire non-identical two point correlation functions near
$T_C$. However, the determinant of the metric tensor or the
geometric invariant objects such as the scalar curvature at zero
quarks chemical potentials indeed remain the same under
$(\frac{\mu}{T})^4$-contributions.

In contrast to the $2$-flavor QCD, the case of 3-flavor QCD shows
that the determinant of the metric tensor, when plotted against the
chemical potentials, has two lines of the minima, and the
corresponding maximum exactly occurs when either one of the
chemical potential reaches the maximum, while the other is held
fixed at the origin. In this case, it is observed that there are
just two distinct bumps of the thermodynamical interactions of
varying heights, instead of the two curves of $2$-flavor QCD.
Since such thermodynamical interactions are local in nature, thus
the height of the observed bumps depend on the domain chosen in
the underlying space of chemical potential which characterizes our
QCD system. We however find in the 3-flavor hot QCD that the shape
of the determinant of the metric tensor appears flatter, and that
the number of bumps is increased under $(\frac{\mu}{T})^4$-contributions,
which in fact indicates that the underlying configuration characterized
as an internal manifold of the chemical potentials is a strongly interacting
statistical system. It turns out that the determinant of the metric tensor,
with the inclusion of thermal fluctuations, acquires two finite
size bumps when plotted against the chemical potentials. In
particular, we observe that there are two bumps of interactions
present when either one of the chemical potentials reaches $ \mu_i
=0.1$, whereas such bumps do not appear in the determinant of the
metric tensor in this range of chemical potentials. It may further
be observed that the strength of the local thermal interactions
becomes diverse with the amplitude of $10^2 GeV^4$ to $10^4 GeV^4$
of the interactions near the $(\mu_1,\mu_2)=(0.1,0.1) GeV$.

\subsection{ Perspective Features}
We have thus provided a very general account of hot QCD
thermodynamic geometries and systematically explored the
underlying thermodynamic space as an intrinsic Riemannian
manifold, and in particular, introduced a covariant geometric
notion for both the case of $2$-flavor and $3$-flavor QCD in the
realism of quasi-particle theories with two distinct chemical
potentials. In both the case of the 2- and 3-favor hot QCDs,
we find that the domain of the thermodynamical interaction gets
shifted towards the origin under the local thermal contributions.
The geometric construction as a covariant intrinsic Riemannian
manifold provides prominent realization of the limiting equilibrium
thermodynamics of the hot QCD near $T_C$. The present investigation
offers an intrinsic geometric exercise of the quasi-particle models
of QCD, which have been described by the quark chemical potentials
with a transition temperature and thus exhibit convincing macroscopic
versus microscopic duality relations via the fermion number density,
as the mixing between different flavors/colors encoded in the
corresponding partition function of effective gauge field theory.

In summary, we have presented a covariant intrinsic
Riemannian geometric perspective of any given analytically
calculable quark number susceptibility tensor, which appears
naturally within an approximately self-consistent resummation of
perturbative QCD. It is worth to mention that the investigations
thus demonstrated are however perturbative in nature, but the same
line of thought may further be applied for the geometric
realization of the underlying quark susceptibility tensors either
in the fabric of lattice QCD or that of non-perturbative QCD, see
for example \cite{hep-lat/0110032}. In particular, it will be
interesting to see whether this kind of approach can be pushed
further, for example to predict geometric properties of chemical
correlations, under sizeable higher order perturbative QCD
contributions, as well as in general to ascertain geometric
features to explore the complete non-perturbative QCD.

%


\vspace{3mm}
\noindent{\bf Acknowledgments:}

VC and BNT acknowledge \textit{``C.S.I.R-New Delhi, India''} for the financial support.

\end{document}